\documentclass[aps,prb,twocolumn,superscriptaddress,showpacs,preprintnumbers]{revtex4-1}
\usepackage[colorlinks,bookmarks=false,citecolor=blue,linkcolor=red,urlcolor=blue]{hyperref}
\usepackage{physics}
\input pdfcolor.tex
\usepackage{colortbl,amsthm,txfonts}
\usepackage{verbatim}
\usepackage{graphicx}
\usepackage{epsfig}
\usepackage{dcolumn}
\usepackage{bm}

\usepackage{amsmath,amssymb}

\usepackage{appendix}

\begin{document}

\title{Fermi-Bose crossover using engineered disorder potential}
\author{Madhuparna Karmakar}
\affiliation{Department of Physics and Nanotechnology, SRM Institute of Science and Technology,   \\ 
Kattankulathur, Chennai 603203, India.}
\email{madhuparna.k@gmail.com}
\author{Rajesh Narayanan}
\affiliation{Department of Physics, Indian Institute of Technology, Madras, \\ Chennai 600036, India.}

\begin{abstract}
We present the first instance of a disorder tuned Fermi-Bose crossover that could be realized in 
superconducting systems. More specifically, harnessing a non perturbative numerical technique we 
analyze the ground state behavior of a two-dimensional attractive Hubbard model subjected to spin 
selective disorder potential. In particular, using spectroscopic properties we provide unambiguous evidence of 
the change in the Fermi surface topology as a function of the disorder, establishing 
incontrovertibly a Fermi-Bose crossover. Interplay of strong correlations and strong disorder brings out the spin selectivity in the properties of this system giving rise to spin selective ``Bose metal/insulator'' phase.
We propose an experimental set-up where such disorder tuned Fermi-Bose crossover could be observed 
in two-dimensional electron gas formed at oxide interface. Finally, we speculate on the possible implication of 
such spin-selective disorder on unravelling signatures of Bose-Fermi cross-over on doped iron chalcogenide superconductors.
\end{abstract}

\date{\today}
\maketitle

\noindent{\bf Introduction:}
The Bardeen-Cooper-Schrieffer (BCS) superfluid phase at weak coupling is continuously connected to the strongly 
coupled phase of the Bose-Einstein condensate (BEC) \cite{randeria_taylor} through a cross-over which 
can be tuned via various control parameters. Of particular interest is 
the cross-over regime at intermediate interactions, where neither a completely fermionic nor a bosonic description of 
the system holds good. Intriguing physics pertaining to the pseudogap phase, preformed Cooper pairs and strange metals are reported to play out in this regime, attributed largely to the short range pair correlations \cite{emery1994,corson1999,li2010,keimer2015}. The advent of ultracold atomic gases made such cross-overs experimentally accessible via the continuous tunability provided by the Feshbach resonance \cite{ketterle_course2006,jin2008,gaebler2010,chin2004,chin_rmp2010,strinati_rmp2008}. 

The Fermi-Bose crossover has been recently realized experimentally in solid state systems such as,  iron based superconductors \cite{kasahara2016,rinott2017,hashimoto2020,kanigel2012}. Furthermore, it has been predicted that transition metal dichalcogenide hetero-structures \cite{Millis_hetero}, could serve as an ideal platform to realize the BEC-BCS cross-over in a topological p-wave superconductor. Apart from the Feshbach  resonance, other novel methods to induce the Fermi-Bose crossover have been proposed which relies on the tuning of the interlayer hopping 
of a prototypical bilayer triangular lattice via strain engineering \cite{loh2016,hazra2021}. 

%%%%%%%%%%%%%%%%%%%%%%%%%%%%%%%%%%%%%%%%%%%%%%%%%%%%%%%%%%%%%%%%%%%%%%%%%%%%%%%%%%%%
\begin{figure}
\begin{center}
\includegraphics[height=8cm,width=8cm,angle=0]{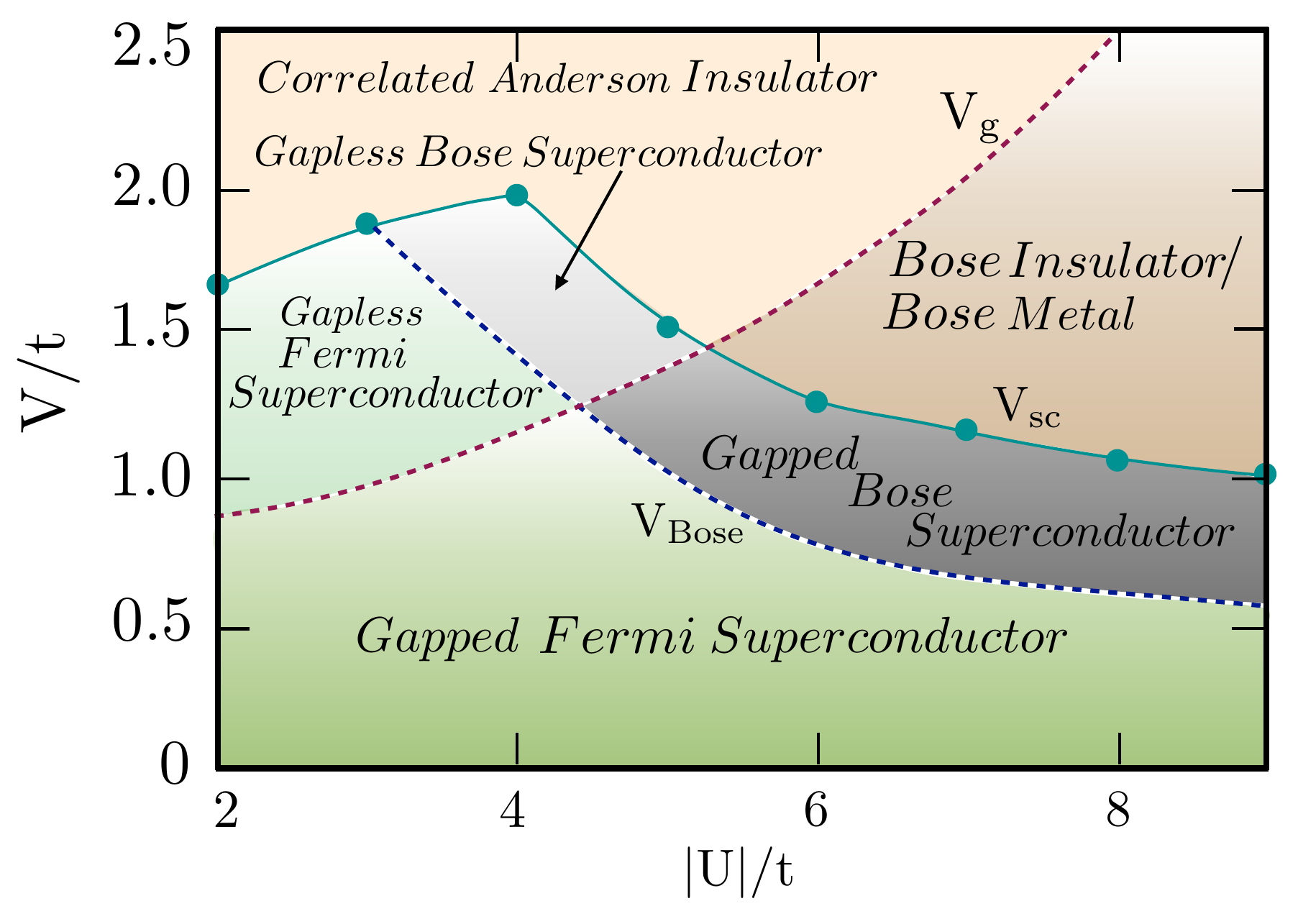}
\caption{Ground state phase diagram in the interaction-disorder ($U-V$)-plane, showing the 
thermodynamic phases viz. (i) gapped superconductor, (ii) gapless superconductor, (iii) spin 
selective Anderson insulator and (iv) spin selective Bose insulator/metal. The dashed curve 
$V_{Bose}$ is the Fermi-Bose crossover scale. The solid curves $V_{g}$ and $V_{sc}$ 
illustrate crossover scales between the gapped and gapless superconductors and superconductor-
insulator transition, respectively. See text for further details.}
\label{fig1}
\end{center}
\end{figure}
%%%%%%%%%%%%%%%%%%%%%%%%%%%%%%%%%%%%%%%%%%%%%%%%%%%%%%%%%%%%%%%%%%%%%%%%%%%%%%%%%%%% 
  
In this work we propose an alternate protocol wherein the Fermi-Bose crossover is achieved via  
the manipulation of the random disorder potential. More specifically, we tune the fermionic dispersion via 
the implementation of a ``spin-selective'' disorder wherein the different spin species are subjected to different 
disorder distributions. Apart from other indicators, we show that the underlying Fermi surface in the BCS superconductor disappears as one tunes through the cross-over into the BEC limit as a function of the spin-selective disorder strength. This  change in the topology of the Fermi surface is a salient feature of the Fermi-Bose cross-over.

The impact of quenched disorder, in particular that of spin independent disorder potential on superconductors, is 
a well investigated subject. These investigations have unravelled, for instance the mechanism behind the localization phenomena \cite{Sacepe2011, Sacepe_natcomm} as well as the nature of the disorder induced superconductor-insulator transition (SIT), in  low dimensional systems  of $d < 3$ \cite{fisher1990quantum, Vojta06, vojta2013phases, Vojta19, Jose_2007, VojtaKotabageHoyos09, ghoshal2001, dubi_nature}.
Furthermore, these studies have revealed the existence of an anomalous metallic phase in certain low dimensional disordered superconductors (see for example Ref.~\cite{kapitulnik2019colloquium}, for a recent review), and also to the observation of exotic disorder induced Griffiths effects in such systems \cite{Aveek, xing2015quantum, shen2016, lewellyn2019infinite, coldea}. In contrast, barring a few exceptions \cite{trivedi_njp,nanguneri2012,RN_MK}, the impact of spin-selective disorder on the superconducting transition has not been so well investigated. 
In particular, Ref.~\cite{RN_MK}, have fully elucidated the phases and the ensuing phase transitions inherent in a spin-selectively disordered two dimensional superconductor, conclusively demonstrating that apart from the usual superconducting and normal phases, the system hosts a breached pair phase (away from the half-filling), wherein superconductivity spatially co-exists with magnetism. 

One such system where a spin-selective disorder tuned Fermi-Bose cross-over can be observed are ultracold atomic gases hosted on an optical lattice \cite{demarco2010,greiner2003,fisher2009,wilczek2004,Mandel_spindep}, wherein the disorder is rendered spin-selective via the introduction of speckle disorder \cite{trivedi_njp,nanguneri2012,Byczuk}. 
Another avenue that we propose where such cross-overs maybe observed is in the two-dimension electron gas 
(2DEG) formed at the interface of oxide heterostructure: for instance, at the oxide interface formed between non magnetic band insulators LaAlO$_{3}$ (LAO) and SrTiO$_{3}$ (STO) 
\cite{ohotomo2004,zhang2019,han2019,xu2017,niu2017,chen2011}. 

Magneto resistance measurements on these systems firmly established the existence of ferromagnetism in the 2DEG \cite{ashoori_nature2010,dikin2011}. Further, more sensitive techniques such as, magnetic mapping by scanning SQUID \cite{bert_nature2011}, magnetic force microscopy \cite{bi2019} etc. established the inhomogeneous spatial coexistence of ferromagnetism with superconducting correlations. Such inhomogeneous local magnetic polarization can be modeled via an inherently random local Zeeman field. An interplay between this random local Zeeman-field and the conventional chemical potential disorder that obtains due to the presence of quenched impurities result in an asymmetric renormalization of the disorder potential seen by the two spin species thus,  leading to spin-selectivity 
of the disorder. Such asymmetric renormalization of the disorder potential has also for instance  been reported in the material Fe$_{1-y}$Co$_{y}$Si wherein the spin-splitting due to the exchange interaction leads to the fact that the 
depth of the potential wells at the Co sites is increased for the majority carriers and decreased for the minority 
carriers \cite{aeppli1}. 

Our key results are:  (i) We propose a novel protocol for tuning the Fermi-Bose cross-over in solid state superconductors using a tailored spin-selective disorder potential. (ii) The cross-over is mapped out in terms of the Fermi surface reconstruction quantified via the spectroscopic signatures. (iii) We demonstrate that the interplay of strong interaction and disorder brings forth a spin-selective Bose
 metal/insulator phase. 
 
\vspace{0.3cm}

\noindent {\bf Model and Methods:}
Our starting point to understand the spin-selective disorder induced Fermi-Bose cross-over is the attractive Hubbard model defined on a square lattice,  described by the Hamiltonian:

\begin{eqnarray}
H & = & -t\sum_{\langle ij \rangle, \sigma}(c_{i\sigma}^{\dagger}c_{j\sigma}
+ c_{j\sigma}^{\dagger}c_{i\sigma})-\mid U \mid \sum_{i}\hat n_{i\uparrow}\hat n_{i\downarrow}
\nonumber \\ && +\sum_{i\sigma}(V_{i\sigma}-\mu)\hat n_{i\sigma}
\label{eq:ham1}
\end{eqnarray}   

\noindent Here, the hopping amplitude is set at $t=1$ for the nearest neighbors, which is the reference energy 
scale of the system. The spin-selectivity comes in making the random $V_{i\sigma}$, spin dependent. 
In our implementation of spin selectivity we have chosen our disorder such 
that $V_{i\uparrow} = 0$ and $V_{i\downarrow} \neq 0 = V$, selected randomly from a box distribution 
of $[-V/2, V/2]$. $\vert U\vert > 0$ is the attractive interaction which gives rise to the $s$-wave superconducting 
pairing. We work in the limit of the half-filled lattice ($n=1$), wherein for the translationally invariant system there 
exists a degeneracy between superconductivity and charge density wave (CDW) orders. Inclusion of disorder immediately lifts this degeneracy and makes the superconducting state energetically favorable over the CDW. 
We tune the global chemical potential $\mu$ to stay in the limit of the half filled lattice. 

As a first step to study Eq.~\ref{eq:ham1} numerically we decompose the interaction term via the Hubbard-Stratonovich (HS) decomposition. Specifically, we introduce a complex $\Delta_{i}(\tau)=\vert \Delta_{i}(\tau)\vert e^{i\theta_{i}(\tau)}$ and a real $\phi_{i}(\tau)$ scalar (bosonic) auxiliary fields at each site, where the former couples to the pairing and the later to the charge channels \cite{hs1,hs2}.
We numerically address this problem by employing a path integral based Monte Carlo technique viz. static path approximated (SPA) quantum Monte Carlo wherein we drop the time dependence of the auxiliary fields and treat them as ``classical''  $\Delta_{i}$ and $\phi_{i}$ \cite{mpk_fflo,evenson1970,dubi_nature}. This is tantamount to retaining only the $\Omega_{n}=0$, Matsubara frequency mode of the fluctuations of the auxiliary fields. The spatial dependence is however, treated exactly by retaining the corresponding spatial fluctuations at all orders, i. e. $\Delta_{i}(\tau) \rightarrow \vert \Delta_{i}\vert e^{i\theta_{i}}$ and $\phi_{i}(\tau) \rightarrow \phi_{i}$, where, $\vert \Delta_{i}\vert$ is the superconducting pairing field amplitude and $\theta_{i}$ is the corresponding phase. These approximations lead to a coupled spin-fermion model with the fermions moving in the spatially fluctuating random background of $\{\Delta_{i}, \phi_{i}\}$, which obeys the Boltzmann distribution, $P\{\Delta_{i}, \phi_{i}\} = \mathrm{Tr}_{c^{\dagger}, c}e^{\beta H_{eff}}$, where $\beta$ is the inverse temperature and $H_{eff}$ is the effective Hamiltonian of the spin-fermion model. For large and random background the trace is computed numerically. The equilibrium configurations of $\{\Delta_{i}, \phi_{i}\}$ are generated via classical Monte Carlo simulation, diagonalizing $H_{eff}$ for each attempted update of $\{\Delta_{i}, \phi_{i}\}$.  The fermionic correlators are computed on these equilibrium configurations of the auxiliary fields. 

The system is characterized using the following indicators: 
$(i)$~superconducting pairing field structure factor ($S({\bf q})=(1/N^{2})\sum_{ij}\langle \Delta_{i}\Delta_{j}^{*}\rangle e^{i{\bf q}.({\bf r}_{i}-{\bf r}_{j})}$), $(ii)$~spin resolved single 
particle density of states (DOS) ($N_{\uparrow}(\omega) = (1/N)\langle \sum_{i, n}\vert u_{n}^{i}\vert^{2}\delta(\omega - E_{n})\rangle$, $N_{\downarrow}(\omega) = (1/N)\langle \sum_{i, n}\vert v_{n}^{i}\vert^{2}\delta(\omega + E_{n})\rangle$) and the associated energy gap $E^{\sigma}_{g}$, $(iii)$~momentum and spin resolved spectral function ($A^{\sigma}({\bf k}, \omega) = (-1/\pi){\rm Im} G^{\sigma}({\bf k}, \omega)$), where, $u_{n}^{i}$ and $v_{n}^{i}$ are 
the Bogoliubov eigenvectors with the eigenvalues $E_{n}$ and $\sigma$ is the spin label. The results presented here corresponds to a system size of $24 \times 24$ and are found to be robust against finite system size effects. The observables are averaged over $10$ disorder realizations and are found to be invariant to the increase in disorder 
realizations. \\

\begin{figure}
\begin{center}
\includegraphics[height=9.0cm,width=9.0cm,angle=0]{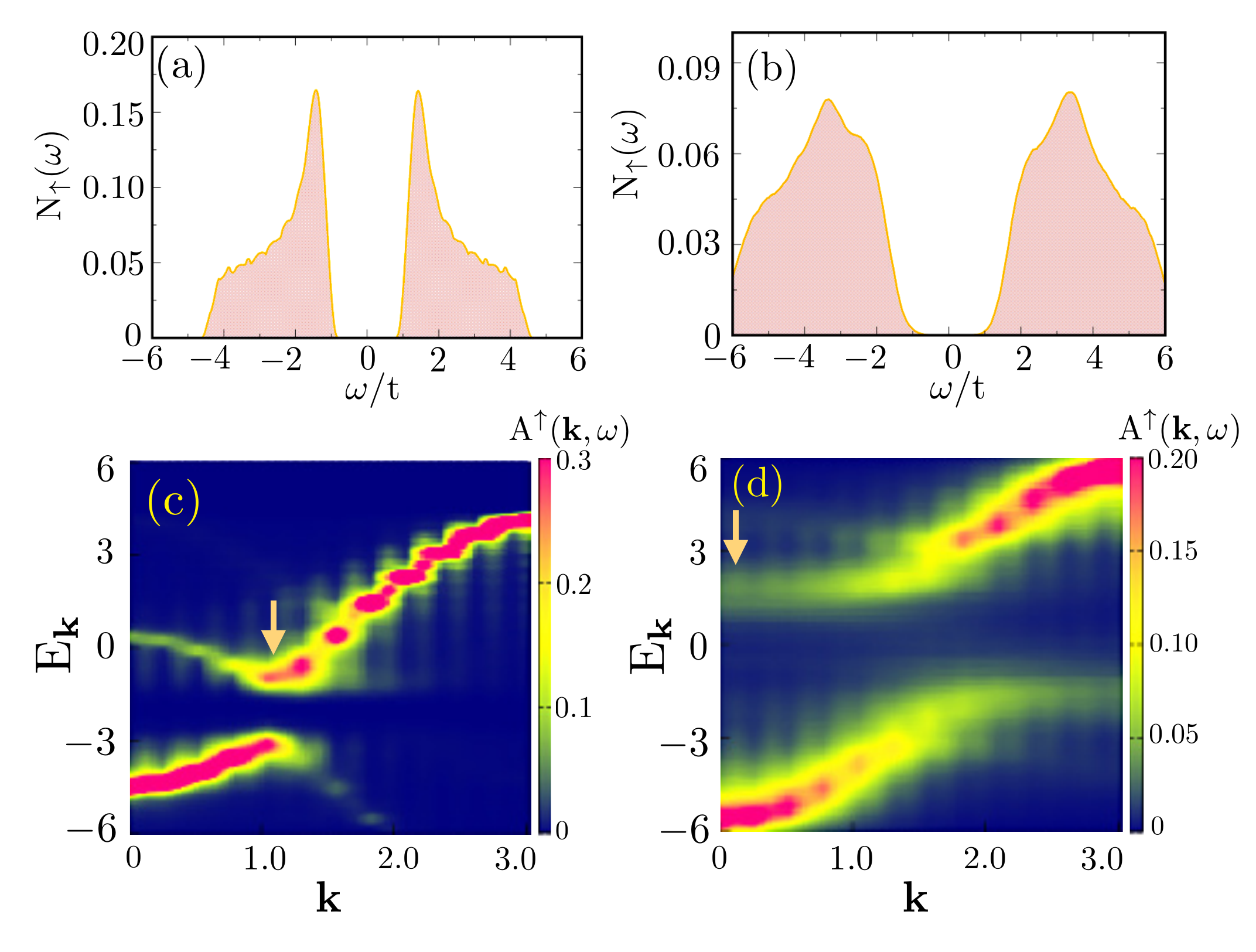}
\caption{Fermi surface topology at the ground state,  at representative $U-V$ cross sections, corresponding 
to fermionic ($U=4t, V=0.25t$) and bosonic ($U=7t, V=t$) superconducting states. 
(a) Single particle DOS for the $\uparrow$-spin species ($N_{\uparrow}(\omega)$) for 
fermionic superconducting phase showing prominent van Hove singularities at the gap 
edges, (b) Single particle DOS for the bosonic superconducting phase showing gap edge discontinuities,
(c) corresponding spectral function ($A^{\uparrow}({\bf k}, \omega)$) showing 
the characteristic back-bending of the Bogoliubov quasi particle spectra with the gap minima 
at ${\bf k} = {\bf k}_{F}$, signifying the presence of underlying Fermi surface. 
(d) corresponding spectral function with flat dispersive branches having gap minima at ${\bf k}=0$, 
signifying the disappearance of the Fermi surface.}
\label{fig2}
\end{center}
\end{figure}

\noindent {\bf The phase diagram:}
Our main results predicated based on the physical observables described above is encapsulated in the ground state phase diagram shown in Fig.~\ref{fig1}: the translationally invariant or clean system clearly is a gapped superconductor. There are 
two cross-over scales labelled as $V_{\rm g}$, and $V_{\rm Bose}$. In the context of this manuscript, one 
of the most relevant scale is $V_{\rm Bose}$, which quantifies the disorder strength where the Fermi superconductor described by a BCS-like state continuously evolves into a strongly coupled BEC state. More specifically, across the cross-over scale set by $V_{\rm Bose}$ the fermionic dispersion spectra, mapped out in terms of the spectral function (see Fig~\ref{fig2}) shows a change in Fermi surface topology: from being downward convex with an energy gap 
minima at a finite wave vector (${\bf k} = {\bf k}_{\rm{F}}$) in the BCS regime, the dispersion branches get modified to being flat, with a gap minima at ${\bf k}=0$ in the BEC regime \cite{loh2016}. This behavior of the spectral function 
showing the BCS-BEC cross-over is accompanied by similar signatures in the single particle DOS: In line with Ref.~\cite{loh2016}, as we tune through the cross-over, the (DOS) sharp van Hove singularities at the superconducting gap edges give way to prominent discontinuities, as the system crosses over from a fermionic to a bosonic state. 

The cross-over scale $V_{\rm g}$ is weakly spin-selective and typifies the disorder strength at which the system evolves from a fully gapped superconductor to a gapless superconductor. The physics of this cross-over is  
very much akin to the effect of magnetic impurities \cite{AG, trivedi_njp,nanguneri2012,RN_MK} on conventional $s$-wave superconductors. The gapped phase is characterized with a finite spectral gap $E_g^{\sigma}$, and a non-vanishing pairing field structure factor $S({\bf q})$. However, in the gapless phase the spectral gap $E^{\sigma}_{g} = 0$, and the $S({\bf q}) \neq 0$.

 This gapless superconductor is fermionic in nature below the cross-over scale $V_{\rm Bose}$, while for $V \ge V_{\rm Bose}$ it evolves into a Bose superconductor.  Finally, the disorder scale $V_{\rm SC}$ encodes the impurity strength 
 at which the transition from the superconducting state to the normal state occurs. Above this scale the pairing field structure factor $S({\bf q}) \rightarrow 0$, indicating the loss of global superconducting phase coherence. This normal state is a gapless correlated Anderson insulator at small interactions which adiabatically morphs into a gapped Bose metal/Insulator with increasing interaction strength.  In what follows,  we discuss the various facets of the data which underpins the construction of the phase diagram shown in Fig.~\ref{fig1}.

 %%%%%%%%%%%%%%%%%%%%%%%%%%%%%%%%%%%%%%%%%%%%%%%%%%%%%%%%%%%%%%%%%%%%%%%%%%%%%%%%%%%%                                  
%%%%%%%%%%%%%%%%%%%%%%%%%%%%%%%%%%%%%%%%%%%%%%%%%%%%%%%%%%%%%%%%%%%%%%%%%%%%%%%%%%%%   

%%%%%%%%%%%%%%%%%%%%%%%%%%%%%%%%%%%%%%%%%%%%%%%%%%%%%%%%%%%%%%%%%%%%%%%%%%%%%%%%
\begin{figure*}
\begin{center}
\includegraphics[height=6cm,width=15cm,angle=0]{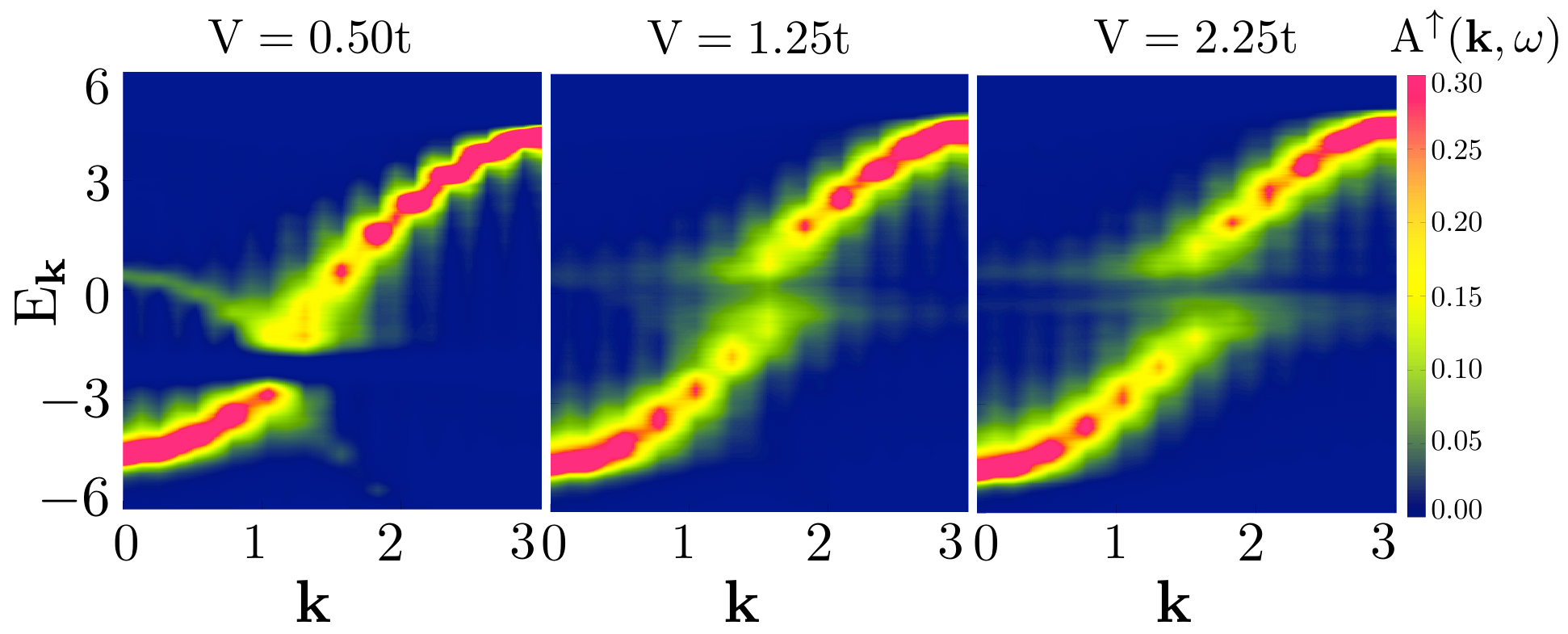}
\caption{Disorder tuned Fermi-Bose crossover as demonstrated via 
single particle spectral function ($A^{\uparrow}(\omega)$) at the ground state 
for $U=4t$. The weak ($V=0.25t$) and strong ($V=2.25t$) limits are characteristic 
to the fermionic and bosonic phases, while at intermediate disorder ($V=1.25t$) 
a (gapless) mixed phase is realized with spin dependent fermionic and bosonic 
characteristics.}
\label{fig3}
\end{center}
\end{figure*}
%%%%%%%%%%%%%%%%%%%%%%%%%%%%%%%%%%%%%%%%%%%%%%%%%%%%%%%%%%%%%%%%%%%%%%%%%%%%%%%%

%%%%%%%%%%%%%%%%%%%%%%%%%%%%%%%%%%%%%%%%%%%%%%%%%%%%%%%%%%%%%%%%%%%%%%%%%%%%%%%%%%%%                                  
\begin{figure}
\begin{center}
\includegraphics[height=6.5cm,width=8.0cm,angle=0]{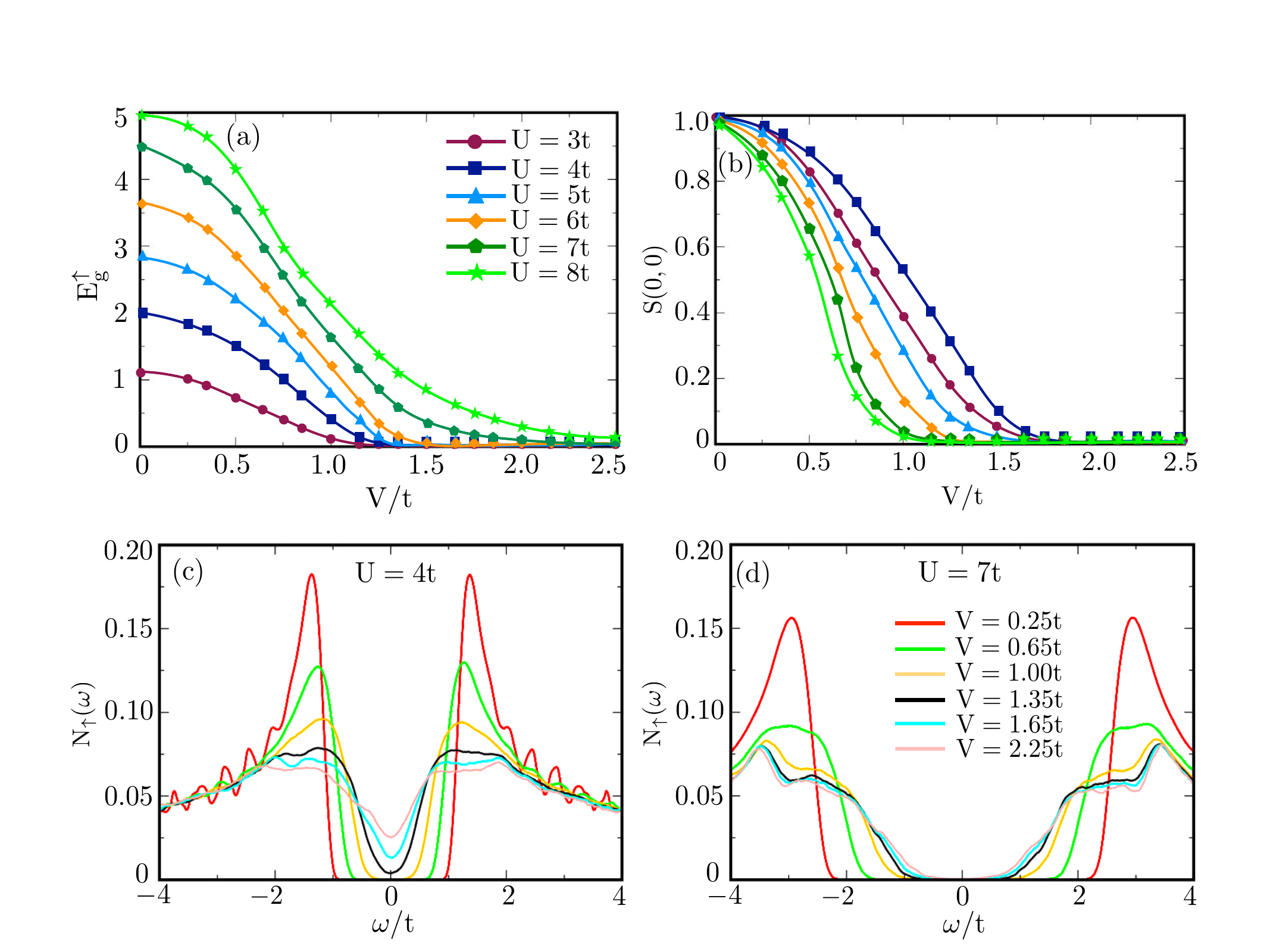}
\caption{Thermodynamic and spectroscopic indicators at the ground state. 
(a) Disorder dependence of the spectral gap ($E_{g}^{\uparrow}$) as determined 
from the corresponding single particle DOS at different interactions. The scale 
$V_{\rm g}$ is set by the gap closing condition $E_{g}^{\uparrow} \rightarrow 0$
(b) Disorder dependent suppression of the superconducting pairing field structure 
factor ($S(0, 0)$) corresponding to the loss of (quasi) long range phase coherence 
($V_{sc}$). (c)-(d) Disorder dependence of the single particle DOS ($N_{\uparrow}(\omega)$) 
at the representative interactions of $U=4t$ and $U=7t$ corresponding to the weak and strong couplings. 
Note the systematic crossover of the gap edges from van Hove singularities at weak disorder to gap edge discontinuities 
at strong disorder, as the system undergoes disorder tuned Fermi-Bose crossover. (see text for further details).}
\label{fig4}
\end{center}
\end{figure}
%%%%%%%%%%%%%%%%%%%%%%%%%%%%%%%%%%%%%%%%%%%%%%%%%%%%%%%%%%%%%%%%%%%%%%%%%%%%%%%%%%%%   

\vspace{0.3cm}

\noindent {\bf Fermi surface reconstruction and Fermi-Bose crossover:}
Here we highlight the Fermi-Bose crossover in detail by using the disorder averaged 
$(i)$ spin resolved single particle DOS ($N_{\sigma}(\omega)$) and $(ii)$ spin resolved spectral function ($A^{\sigma}({\bf k}, \omega)$), shown in Fig.~\ref{fig2}. In particular, Fig.~\ref{fig2} compares the representative behavior of the spin-resolved DOS $N_{\uparrow}(\omega)$ and the spin-resolved spectral function $A^{\uparrow}({\bf k}, \omega)$ 
in the Fermi superconducting phase (for example at,  $U=4t$, $V=0.25t$), (see Fig.~\ref{fig2}(a) 
and \ref{fig2}(c) respectively), with their behavior in the Bose superconducting phase, (see Fig.~\ref{fig2}(b) and Fig.~\ref{fig2}(d)) (exemplified by the situation at $U=7t$, $V=t$). The Fermi superconducting phase is characterized by a robust spectral gap at the Fermi level with sharp van Hove singularities (inverse square root divergence) at the gap edges, as shown in Fig.~ \ref{fig2}(a). The corresponding spectral function shown in Fig.~\ref{fig2}(c) contains the signatures of the usual Bogoliubov spectrum comprising of the dispersion branches $\pm E_{\bf k}$ with weights 
$u_{\bf k}^{2}$ and $v_{\bf k}^{2}$ ($u_{\bf k}$ and $v_{\bf k}$ are the eigenvectors corresponding to the energy eigenvalues $\pm E_{\bf k}$). The spectral function shows that the minimum energy gap occurs on a 
contour at a finite wave vector ${\bf k} = {\bf k}_{\rm F}$, indicating the presence of an underlying Fermi surface. 
   
On the other hand, the spectral function corresponding to the Bose superconducting phase (Figure \ref{fig2}(d)) comprises of a single point (instead of a locus) at the wave vector ${\bf k}=0$, where the gap minima occurs, 
indicating the absence of a Fermi surface. The corresponding single particle DOS, shown in Figure \ref{fig2}(b), 
no longer hosts van Hove singularities, and rather exhibits discontinuity or kink at the gap edges. 
This somewhat smoothened (due to the effect of quenched disorder), discontinuity is very reminiscent of the
 jump in the DOS predicted to occur in the parameter window where a bosonic description obtains Ref.~\cite{loh2016}. 
The system thus undergoes a disorder tuned Fermi-Bose crossover, across which the Fermi surface undergoes 
a change in its topology. The spectroscopic signatures characteristic of this crossover can be experimentally 
observed via radio frequency (rf) and angle resolved photo emission spectroscopy (ARPES) measurements. 

The systematic evolution of the Fermi surface topology with the disorder potential at $U=4t$ is shown in Figure 
\ref{fig3}, in terms of the fermionic spectral function $A^{\uparrow}({\bf k}, \omega)$. As a function of increasing 
disorder the spectrum undergoes cross-over from being downward convex (minima at ${\bf k} = {\bf k}_{\rm F}$) to flat 
(minima at ${\bf k} = 0$) as the underlying Fermi surface vanishes. At intermediate disorder potential a 
gapless superconducting state is realized; for the corresponding dispersion spectra neither a completely 
fermionic nor a completely bosonic description holds good. The system exhibits a spin-selective behavior such 
that, the $\uparrow$-spin species show signatures of bosonic correlations in its spectroscopic properties while the $\downarrow$-spin species behaves akin to a fermionic system. We observe that the disorder tuned evolution of the Fermi surface topology is in agreement with the one reported recently in ARPES measurements across the 
Fermi-Bose crossover in doped Fe-based superconductor, FeSe$_{1-x}$S$_{x}$ \cite{hashimoto2020}.

{\bf The energy scales $V_{\rm g}$ and $V_{\rm SC}$}:
We now use the disorder dependence of $E^{\uparrow}_{g}$ ($E^{\downarrow}_{g}$ behaves similarly) 
and $S({\bf q}=0)$ at different interactions (as shown in Fig.~\ref{fig4}(a) and \ref{fig4}(b), respectively), to capture 
the scales $V_{\rm g}$, and $V_{\rm SC}$,  respectively. In Fig.~\ref{fig4}(a), we plot the behavior of 
$E^{\uparrow}_{g}$ as a function of the disorder $V/t$, for a range of interactions. 
The energy scale $V_{\rm g}$ is gleaned from Fig.~\ref{fig4}(a) as the disorder strength (for a particular 
value of $U/t$) at which the $E^{\uparrow}_{g} \rightarrow 0$. The $E^{\uparrow}_{g}$ is tied to the pairing field amplitude ($\vert \Delta_{i}\vert$) which monotonically increases with increasing $U/t$ and this monotonic increase 
is reflected on the scale $V_{\rm g}$, which also shows concomitant monotonic increase with $U/t$, (see Fig.~\ref{fig1}).

To extract the energy scale $V_{\rm SC}$, we lay recourse to the pairing field
structure factor $S({\bf q})$. In Fig.~\ref{fig4}(b), we depict the behavior of $S({\bf q})$ as a function of 
$V/t$ for various interaction strength $U/t$. The scale $V_{\rm SC}$  is captured by the value of disorder strength at which the pairing field structure factor vanishes for each value of $U/t$.  From Fig.~\ref{fig4}(b), we immediately see that 
$S({\bf q})$ and consequently $V_{\rm SC}$ display a non-monotonic behavior as a function of interaction strength. This non-monotonicity of the $S({\bf q})$, and subsequently $V_{\rm SC}$, (especially in comparison with  with increasing $E^{\uparrow}_{g}$), as a function of $U/t$ is predicated on the different origins that control the loss of superconducting order in the weak coupling and the intermediate and strong coupling limits: 
In the weak-coupling limit, the loss of the superconducting order is tied to the suppression of the pairing field amplitude, (thus implying that $E^{\uparrow}_{g}$, is a suitable indicator), whereas at intermediate and strong interactions it is the phase fluctuations that are responsible for the suppression of the (quasi) long range order. 
%%%%%%%%%%%%%%%%%%%%%%%%%%%%%%%%%%%%%%%%%%%%%%%%%%%%%%%%%%%%%%%%%%%%%%%%%%%%%%%%%%%%           
\begin{figure*}
\begin{center}
\includegraphics[height=9.0cm,width=16.0cm,angle=0]{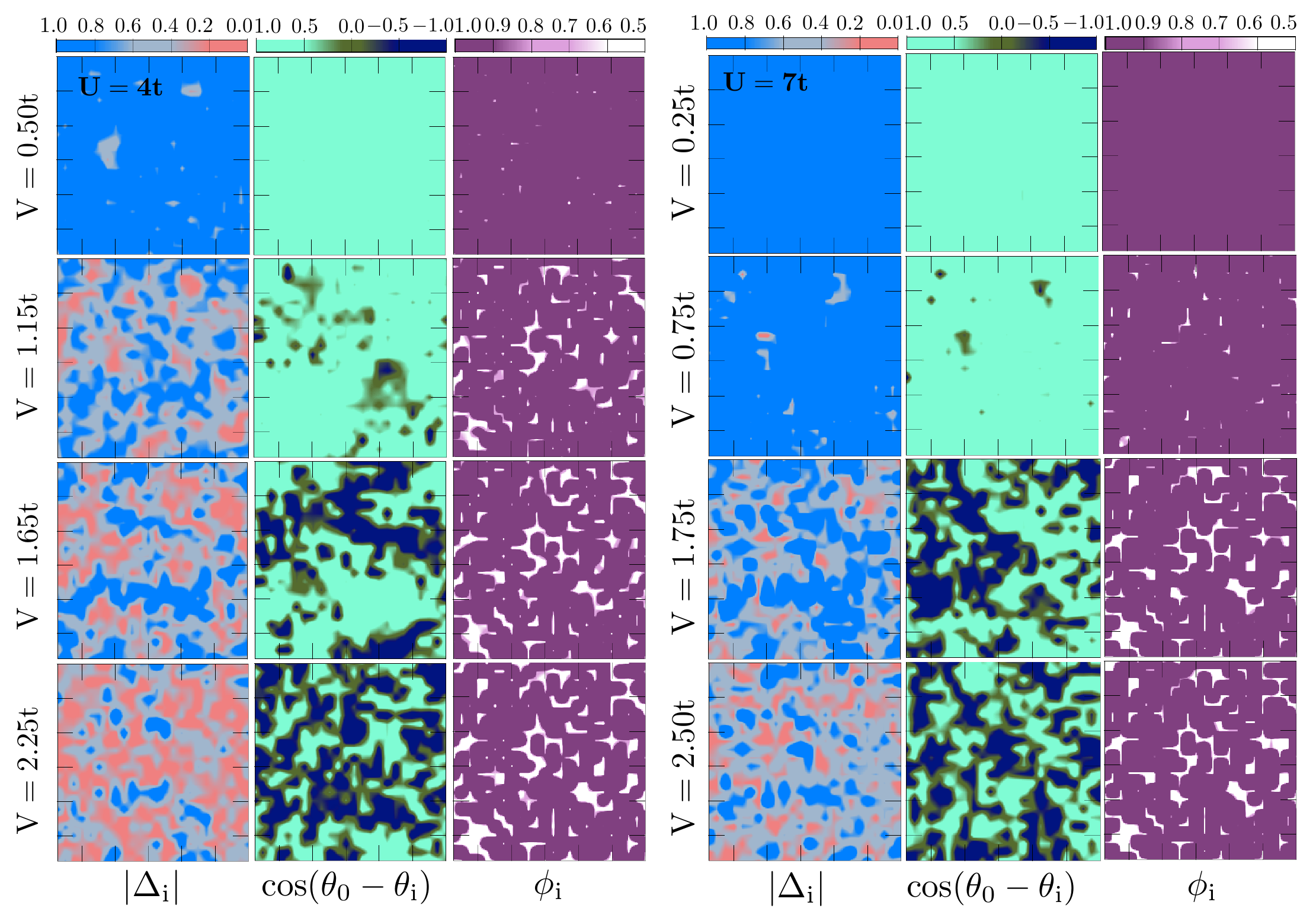}
\caption{Real space maps corresponding to the superconducting pairing field amplitude ($\vert \Delta_{i}\vert$), 
superconducting phase correlation between a reference site ($0$) and any other lattice site ($\cos(\theta_{0}-\theta_{i})$) and the charge field ($\phi_{i}$), at selected interaction-disorder ($U-V$) cross sections. The spatial maps characterize the excursion of the system from a gapped Fermi superconductor $\rightarrow$ gapless Fermi superconductor $\rightarrow$ gapless Bose superconductor $\rightarrow$ correlated Anderson insulator, as a function of disorder in  the weak coupling regime ($U=4t$). In the representative strong coupling regime ($U=7t$), the system follows the trajectory of a gapped Fermi superconductor $\rightarrow$ gapped Bose superconductor $\rightarrow$ spin selective Bose metal/insulator $\rightarrow$ correlated Anderson insulator, as the function of disorder.}
\label{fig6}
\end{center}
\end{figure*}
%%%%%%%%%%%%%%%%%%%%%%%%%%%%%%%%%%%%%%%%%%%%%%%%%%%%%%%%%%%%%%%%%%%%%%%%%%%%%%%%%%%

The behavior of the single particle DOS ($N_{\uparrow}(\omega)$) across the disorder tuned Fermi-Bose 
crossover is shown in Figure \ref{fig4}(c) and \ref{fig4}(d), at two different interactions $U=4t$ and 
$U=7t$, respectively. This behavior of the DOS exhibited in Fig.~\ref{fig4}(c) should be read in conjunction with Fig.~\ref{fig6} which depicts the real space signatures of the system across the selected $U-V$ cross sections 
in terms of the pairing field amplitude ($\vert \Delta_{i}\vert$) and pairing field phase coherence between a reference site and all other sites of the lattice ($\cos(\theta_{0}-\theta_{i})$).  For the sake of completeness the spatial maps pertaining to the bosonic auxiliary field $\phi_{i}$ coupled to the charge channel is also presented.
At $U=4t$ the system is a gapped Fermi superconductor upto $V_{g} \sim 1.10t$, as attested by the robust spectral 
gap at the Fermi level with inverse square root singularities at the gap edges,  as shown in Figure \ref{fig4}(c). In the gapped Fermi superconductor regime, (as typified for the case $U=4t$, $V=0.5t$ in Fig.~\ref{fig6}), for all disorder strengths below the scale $V_{Bose}$,  the system corresponds to a (quasi) long range phase cohered uniform Fermi superconductor characterized by a robust and uniform pairing field amplitude.

Over the regime  $1.10t \lesssim V < V_{Bose} \sim 1.25t$ there is finite spectral weight at the Fermi 
level accompanied by the suppression of the coherence peaks at the gap edges, characteristic of the 
gapless Fermi superconductor. The real space signature of this gapless Fermi superconductor (at the prototypical value of $U=4t$, $V=1.15t$), indicate that the corresponding phase coherence is still (quasi) long ranged but the pairing field amplitude is spatially fragmented into "correlated" islands of suppressed $\vert \Delta_{i}\vert$.
At $V_{Bose} \sim 1.3t$, discontinuity appears at the gap edges, indicating that the underlying phase is now a Bose superconductor.  
The gapless superconductor is characterized by a finite spectral weight at the Fermi level, and by 
a small but finite value for the superconducting structure factor $S(0,0)$,  shown in Fig.~\ref{fig4}(b).
This phase is characterized by a real space picture, (see Fig.~\ref{fig6} for $U=4t$, $V=1.65t$) wherein one sees a pronounced suppression of $\vert \Delta_{i}\vert$. Furthermore, the loss of phase coherence in the gapless Bosonic superconductor is also affirmed by the spatial maps for the phase correlations shown in Fig.~\ref{fig6} (for $U=4t$, $V=1.65t$). 

For $V \gtrsim V_{sc}$ a correlated Anderson insulator obtains characterized by a gapless spectra, accompanied by large transfer of spectral weight away from the Fermi level and absence of any phase coherence peak at the gap edges. The gapless Bose superconductor and the correlated Anderson insulator are distinguished from each other by the presence of (quasi) long range superconducting phase coherence, in the former and its absence in the later phase. The real space signature of the correlated Anderson Insulator, ( typified by the case at $U=4t, V=2.25t$ in Fig.~\ref{fig6}) show the lack superconducting correlations both in terms of the pairing field amplitude and phase, as observed via the corresponding spatial maps.

Now,  at strong coupling for instance at $U=7t$, the system is a gapped Fermi superconductor up to $V_{Bose} \lesssim 0.65t$, shows spatial signatures that corresponds to a robust (quasi)-long range phase cohered uniform Fermi superconductor characterized by a spatially homogenous  pairing field amplitude as seen for the case $U=7t$, $V=0.25t$, akin to the weak coupling scenario discussed above. At strong coupling, in  the regime $V_{Bose} \le V < V_{sc} \sim 1.10t$, the system is a gapped Bose superconductor: The DOS in this regime, (see Fig.~\ref{fig4}(d)), clearly exhibits a well developed gap at the Fermi level, and the appearance of discontinuity at the gap edges. This gapped Bose superconductor (exemplified by  $U=7t, V=0.75t$ in Fig.~\ref{fig6}), is characterized with a real space map that shows  a robust $\vert \Delta_{i}\vert$ and (quasi) long range phase coherence.
Over the regime $V_{sc} \lesssim V < V_{g} \sim 2t$, the system is a spin-selective Bose insulator 
or metal with a robust spectral gap at the Fermi level and absence of superconducting phase coherence. 

Fig.~\ref{fig6} (for the case $U=7t, V=1.75t$), illustrates the prototypical behavior of the gap and the phase correlations in such a spin-selective Bose metal/insulator: this state is phase uncorrelated but hosts fairly large amplitude $\vert \Delta_{i}\vert$, which suggests that the system is characterized by a random  distribution of the 
pairing field with a suppressed magnitude, i. e. $\vert \Delta_{i}\vert$ are large but randomly distributed. 
The strong coupling regime, for $V>V_{\rm g}$ as evidenced for the typical case ($U=7t$, $V=2.50t$) is a correlated Anderson insulator. Pairing field amplitudes are significantly suppressed in this regime 
 akin to that of the weak coupling, represented by $U=4t, V=2.25t$.
 %Akin to $U=7t, V=1.75t$, this phase with bosonic correlation comprises of large amplitude $\vert \Delta_{i}\vert$. 
 
The fermionic number density remains pinned at the half filling ($n=1$) across the $U-V$ space as attested by the largely uniform distribution of $\phi_{i}$, barring the strong disorder regime where weak local fluctuations are observed in the spatial distribution of $\phi_{i}$.
Real space probes for instance,  STM would be able to discern the changes in the DOS as we tune through the Bose-Fermi cross-over \cite{Ketterman_Xover}. 
%%%%%%%%%%%%%%%%%%%%%%%%%%%%%%%%%%%%%%%%%%%%%%%%%%%%%%%%%%%
\begin{figure}
\begin{center}
\includegraphics[height=6.5cm,width=8.0cm,angle=0]{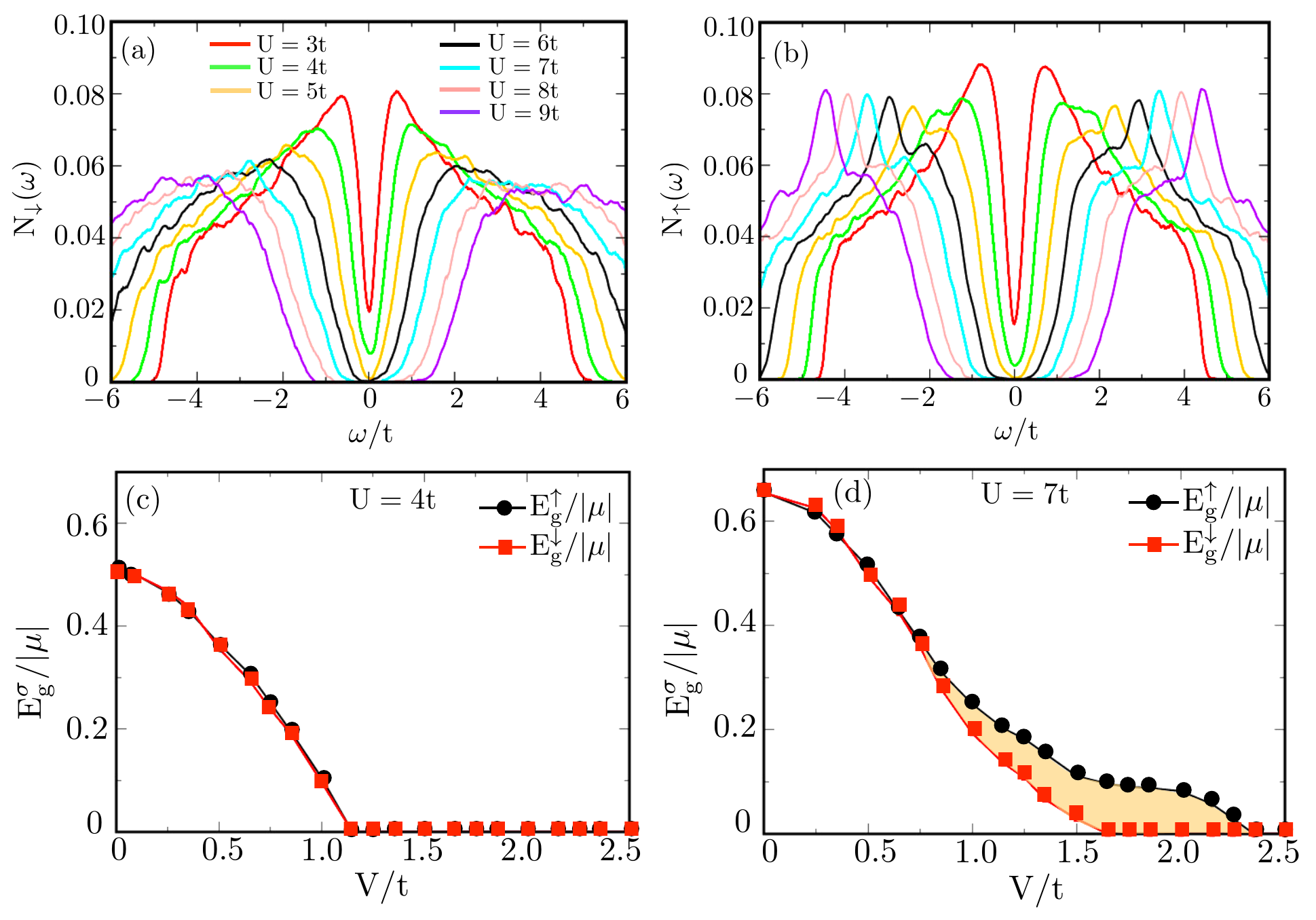}
\caption{Spin-selective behavior of the DOS at $V=1.35t$ as a function of the interaction 
strength: In the strong 
coupling regime $N_{\downarrow}(\omega)$ (see Fig.~\ref{fig4_alt}(a)), undergoes large transfer of spectral weight 
away from the Fermi level indicating a Bose insulator phase. In comparison, the $N_{\uparrow}(\omega)$ as seen in  
Fig.~\ref{fig4_alt}(b) shows prominent gap edge discontinuities. (c)-(d) Disorder dependence of the ratio between the spin selective energy gap at the Fermi level ($E_{g}^{\sigma}$) and the global chemical potential ($\mu$) adjusted to the Fermi energy, as obtained from our numerical calculations at $U = 4t$ and $U = 7t$ respectively. At weak coupling both the spin species behaves identically, however the spin selective dependence on the disorder potential becomes prominent at strong coupling. The shaded region in panel (d) corresponds to the regime where only one of the spin species is rendered gapless by disorder.}
\label{fig4_alt}
\end{center}
\end{figure}
%%%%%%%%%%%%%%%%%%%%%%%%%%%%%%%%%%%%%%%%%%%%%%%%%%%%%%%%%%%

\vspace{0.3cm}

\noindent{\bf Spin selectivity:}
In Fig.~\ref{fig4_alt}(a) and Fig.~\ref{fig4_alt}(b) we contrast the spin resolved single particle DOS for different 
interactions at a selected disorder potential of $V=1.35t$. At this choice of the disorder potential the system 
is a gapless Fermi superconductor for $U=3t$ and a gapless Bose superconductor at $U=4t$ and $U=5t$ 
{(see Fig.~\ref{fig4}(b) for details). Over the regime $U \gtrsim 6t$ a spin selective Bose 
metal/insulator state is realized. We understand these phases based on the $N_{\sigma}(\omega)$ 
as follows: at $U=3t$ there is finite spectral weight at the Fermi level in the single particle DOS, along with prominent phase coherence peaks at the gap edges, as is expected from a gapless Fermi 
superconductor. The spectrum continues to be gapless upto $U=5t$ following which the spectrum contains broader peaks with weak signatures of discontinuity, indicating a cross-over into a gapless Bose superconducting state.  The spectral gap opens up at $U \ge 6t$ and the discontinuity at the gap edges get progressively pronounced spin-selectively: with increasing interaction for the $\uparrow$-spin species the 
discontinuities at the gap edges become more pronounced. In contrast,  for the $\downarrow$-spin species there is progressively larger transfer of weight away from the Fermi level with the corresponding gap edge discontinuities becoming more rounded. The spin dependent dichotomy in the behavior of the DOS can be easily ascribed to the fact that the $\downarrow$-spin species sees the disorder directly thus rounding the discontinuities at the gap edges very strongly, (refer Fig.~\ref{fig4_alt}(a)). In contrast, the $\uparrow$-spin species sees a renormalized disorder via residual interactions and thus sees a weaker disorder,  as a consequence of which the discontinuities at the gap edges in the $N_{\uparrow}(\omega)$ are more prominent, (see Fig.~\ref{fig4_alt}(b)). 
%%%%%%%%%%%%%%%%%%%%%%%%%%%%%%%%%%%%%%%%%%%%%%%%%%%%%%%%%%%%%%%%%%%%%%%%%%%%%%%%%%%%%%%%
This spin-selectiveness is once again brought to the fore in Fig.~\ref{fig4_alt}(c) and Fig.~\ref{fig4_alt}(d) which depicts 
the disorder dependence of the ratio of the spin-selective gap to the global chemical potential (set to the Fermi energy), for two prototypical values at,  $U= 4t$, and $U=7t$. In the weak coupling regime, ($U=4t$), as seen in Fig.~\ref{fig4_alt}(c) the ratio of the superconducting gap to the global chemical potential $E_{\rm g}^\sigma/|\mu|$, behaves identically for both the spin species. However, in the strong coupling regime for instance,  at the prototypical value $U=7t$, the same ratio of $E_{\rm g}^\sigma/|\mu|$ shows asymmetric renormalization as a function of the disorder strength, (see Fig.~\ref{fig4_alt}(d)): The superconducting gap completely collapses for the $\downarrow$-spin fermions as a function of the disorder strength whilst the superconducting gap is still robust for the $\uparrow$-spin species.

We now turn our attention to the nature of the non superconducting state that obtains at large values of $U$ above the scale set by $V_{\rm sc}$: the anomalous spectrum of the $\uparrow$-spin species in this parameter range characterized by $(i)$ a spectral gap at the Fermi level, $(ii)$ pronounced discontinuity at the gap edges and $(iii)$ phase uncorrelated puddles of superconducting pairing field with large amplitude (shown in Fig.\ref{fig6})  \cite{phillips2005,pashupathy2015} leads us to speculate that for $V \ge V_{\rm sc}$, for large values of interaction strength one might stabilize a spin-selective Bose metal or insulator.
 The $\downarrow$-spin species in contrast loses the bosonic correlations faster owing to the direct influence of the disorder potential. We note that the exact nature of this spin selective Bose metal/insulator regime can be ascertained only via the computation of the relevant transport signatures,  a task which is beyond the scope of this work.

\vspace{0.3cm}

\noindent {\bf Discussion and conclusions:}
 In conclusion, we have demonstrated that suitably engineered random disorder potential can serve 
as a tuning parameter for the Fermi-Bose crossover in 2D superconductor. Based on the thermodynamic 
and spectroscopic signatures as determined via a non perturbative numerical approach to the problem 
we have established the Fermi surface reconstruction as the system undergoes disorder tuned Fermi-Bose 
crossover. Potential candidates in solid state systems where such Fermi-Bose cross-over can be realized are the
the iron chalcogenides \cite{shimozima2017, miao2015}. These materials  have the ratio $E_{g}/\epsilon_{F} \sim 1$ which could be renormalized across the cross-over regime by chemical doping: for example in FeSe$_{1-x}$S$_{x}$ \cite{hashimoto2020}, it has been observed that substituting Se by the isovalent S in FeSe results in significant suppression of the $E_{g}/\epsilon_{F}$ ratio, without altering the carrier concentration in the material. This 
suppression of the $E_{g}/\epsilon_{F}$ ratio is very qualitatively reminiscent of the suppression seen for 
example in Fig.~\ref{fig4_alt}(c), and Fig.~\ref{fig4_alt}(d). However, the single orbital model used in our work 
does not take into account the multiorbital band structure of real FeSe superconductors. It is expected that a 
multiorbital approach to the problem will bring in complexities arising out of the interplay between the inter and 
intra-band superconducting pairing. It has been recently demonstrated that for a multiorbital model the pseudogap phase is orbital selective and gives rise to rich BCS-BEC cross-over physics \cite{tajima2019}. Our work opens up a tantalizing possibility that the BCS-BEC cross-over physics in iron chalcogenides could plausibly occur when different orbitals see different disorder: a sort of orbital selective counterpart of the spin-selective disorder seen in this manuscript. 

We next turn our attention to some open problems with regards to the effect of spin-selective disorder 
in superconducting systems: for instance, an open problem is to calculate the transport signatures so as to  
completely characterize the nature of the normal state (i. e. Bose metal/insulator regime).  In a similar vein, the 
impact of thermal fluctuations in altering the ground state properties is a problem that is being currently pursued \cite{MKRN}.  This is particularly relevant as thermal fluctuations induced spatial fragmentation of the superconducting state into isolated phase uncorrelated islands with large superconducting pairing amplitude could give rise to the pseudogap phase at $T \neq 0$. Recently it has been argued that the finite temperature pseudogap phase is in fact a Bose metal \cite{trivedi2020}. In a similar spirit the disorder induced spatial fragmentation of the superconducting state can be thought of as the origin of the low temperature Bose metal phase.      

\textit{Acknowledgements:}
The authors acknowledge the use of high performance computing cluster Aqua, at IIT, Madras, India. R.N. 
acknowledge funding from the Center for Quantum Information Theory in Matter and Spacetime, IIT Madras and from the Department of Science and Technology, Govt. of India, under Grant No. DST/ICPS/QuST/Theme-3/2019/Q69. 
We thank Prof. G. Baskaran (IIT-Madras) for valuable discussions. We also thank Dr. Prabuddha Chakraborty (ISI-Bangalore), for a collaboration in the early part of this project.
 
\bibliographystyle{apsrev4-1}
\bibliography{spinres_rev1.bib}
\end{document}